\documentclass[useAMS,usenatbib]{mn2e}
\usepackage[dvips]{graphicx}
\newcommand{\bz}{$\langle B_z \rangle$}
\newcommand{\nz}{$\langle N_z \rangle$}
\newcommand{\vsini}{$v \sin i$}
\newcommand{\kms}{km\,s$^{-1}$}

\title[The radial crossover signature of $\xi^1$ CMa]{The pulsationally modulated radial crossover signature of the slowly rotating magnetic B-type star $\xi^1$ CMa\thanks{Based on observations obtained with ESPaDOnS at the Canada-France-Hawaii Telescope (CFHT), which is operated by the National Research Council of Canada, the Institut National des Sciences de l'Univers of the Centre National de la Recherche Scientifique of France, and the University of Hawaii.}}
\author[M. Shultz]
{M.\ Shultz$^{1}$\thanks{E-mail: matthew.shultz@physics.uu.se},
O.\ Kochukhov$^{1}$,
G.\ A. Wade$^{2}$,
Th.\ Rivinius$^{3}$\\
$^1$Department of Physics and Astronomy, Uppsala University, Box 516, Uppsala 75120, Sweden \\
$^2$Department of Physics and Space Science, Royal Military College of Canada, Kingston, Ontario K7K 7B4, Canada\\
$^3$ESO - European Organisation for Astronomical Research in the Southern Hemisphere, Casilla 19001, Santiago 19, Chile\\
}
\begin{document}

\date{}

\pagerange{\pageref{firstpage}--\pageref{lastpage}} \pubyear{2018}

\maketitle

\label{firstpage}

\begin{abstract}
We report the latest set of spectropolarimetric observations of the magnetic $\beta$ Cep star $\xi^1$ CMa. The new observations confirm the long-period model of Shultz et al.\ (2017), who proposed a rotational period of about 30 years and predicted that in 2018 the star should pass through a magnetic null. In perfect agreement with this projection, all longitudinal magnetic field $\langle B_z \rangle$ measurements are close to 0 G. Remarkably, individual Stokes $V$ profiles all display a crossover signature, which is consistent with $\langle B_z \rangle \sim 0$ but is {\em not} expected when $v\sin{i} \sim 0$. The crossover signatures furthermore exhibit pulsationally modulated amplitude and sign variations. We show that these unexpected phenomena can all be explained by a `radial crossover' effect related to the star's radial pulsations, together with an important deviation of the global field topology from a purely dipolar structure, which we explore via a dipole+quadrupole configuration as the simplest non-dipolar field. 
\end{abstract}


\section{Introduction}

The magnetic $\beta$ Cep star $\xi^1$ CMa (HD 46328, B0.5 IV) occupies a unique place amongst the magnetic B-type stars. It has the strongest magnetic field of any $\beta$ Cep star; it is the only magnetic star with magnetospheric emission modulated by pulsation as well as rotation, in both X-rays and H$\alpha$ \citep{2014NatCo...5E4024O,2017MNRAS.471.2286S}; and it is the coolest magnetic star with magnetospheric emission profiles consistent with an origin in a dynamical rather than a centrifugal magnetosphere \citep{2017MNRAS.471.2286S}. Since dynamical magnetospheres are more typically detectable in H$\alpha$ in magnetic O-type stars \citep{petit2013}, this makes $\xi^1$ CMa an important transitional object. 

\cite{2017MNRAS.471.2286S} (hereafter S17) showed that $\xi^1$ CMa has by far the longest known period of any magnetic B-type star, at least 30 years. At present, modern high-resolution magnetic data covers no more than about 20\% of the rotation period (two additional MuSiCoS observations acquired in 2000 extend this to about 67\%). Completion of phase coverage requires continuing observation with an approximately annual cadence. In this Letter, we report the results of the most recent season of high-resolution ESPaDOnS spectropolarimetry, obtained for the first time close to a magnetic null. In \S~\ref{sec:obs} we describe the observations and magnetic measurements. The results of our magnetic analysis are presented in \S~\ref{sec:results}. In \S~\ref{sec:discussion} we analyse and interpret the results, and our findings are summarized in \S~\ref{sec:conclusions}.

\section{Observations \& Magnetometry}\label{sec:obs}

\begin{table}
\centering
\caption[\bz~measurements]{Table of recently acquired RV and \bz~measurements. Pulsation phases were calculated using the non-linear ephemeris given by S17. ``DF'' is the detection flag (described in more detail in the text).}
\resizebox{8.5 cm}{!}{
\begin{tabular}{llrrlrl}
\hline
\hline
        &           &        & \multicolumn{2}{c}{Stokes $V$} & \multicolumn{2}{c}{Null} \\
HJD -   & Pulsation & RV     & \bz & DF & \nz & DF \\
2458000 & Phase     & (\kms) & (G) &    & (G) &    \\
\hline
148.75483 & 0.36201 & 11.5 & 22$\pm$10 &  DD &  9$\pm$10 &  ND \\
148.97306 & 0.40330 &  8.8 &  8$\pm$ 8 &  DD &  0$\pm$ 8 &  ND \\
150.80051 & 0.12295 & 34.9 & -4$\pm$11 &  DD & -2$\pm$11 &  ND \\
153.85673 & 0.70567 & 19.5 & 11$\pm$ 8 &  DD & -5$\pm$ 8 &  ND \\
154.70953 & 0.77480 & 26.8 &  2$\pm$12 &  DD & 12$\pm$12 &  ND \\
154.91499 & 0.75515 & 24.3 &  8$\pm$ 8 &  DD & -4$\pm$ 8 &  ND \\
156.76559 & 0.58526 &  9.7 &  8$\pm$12 &  MD & -3$\pm$12 &  ND \\
156.91589 & 0.30242 & 15.9 & 10$\pm$11 &  MD & -6$\pm$11 &  ND \\
\hline
\hline
\end{tabular}
}
\label{obs_log}
\end{table}

   \begin{figure}
   \centering
   \includegraphics[width=\hsize]{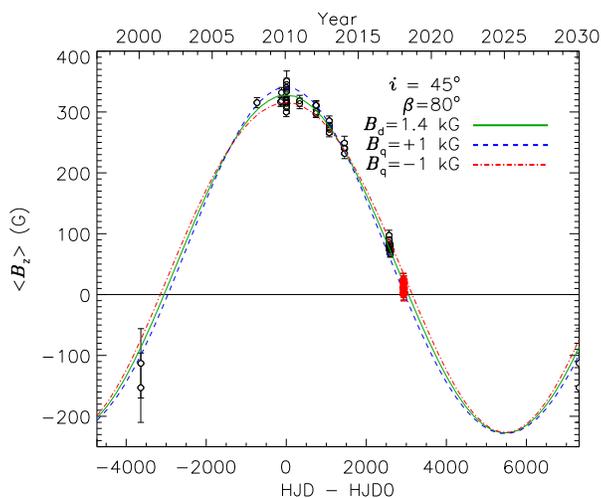} 
      \caption[]{\bz~as a function of time. New measurements are shown as filled red circles. The solid green curve shows the best-fitting dipolar model assuming a 30-year rotation period, with ${\rm HJD0} = 2455219$. The dashed blue and dot-dashed red curves show the same model with $B_{\rm q} = \pm 1~{\rm kG}$. The three \bz~curves are essentially indistinguishable.}
         \label{xi1cma_bz_bq}
   \end{figure}

   \begin{figure*}
   \centering
	\begin{tabular}{cc}
   \includegraphics[width=9cm]{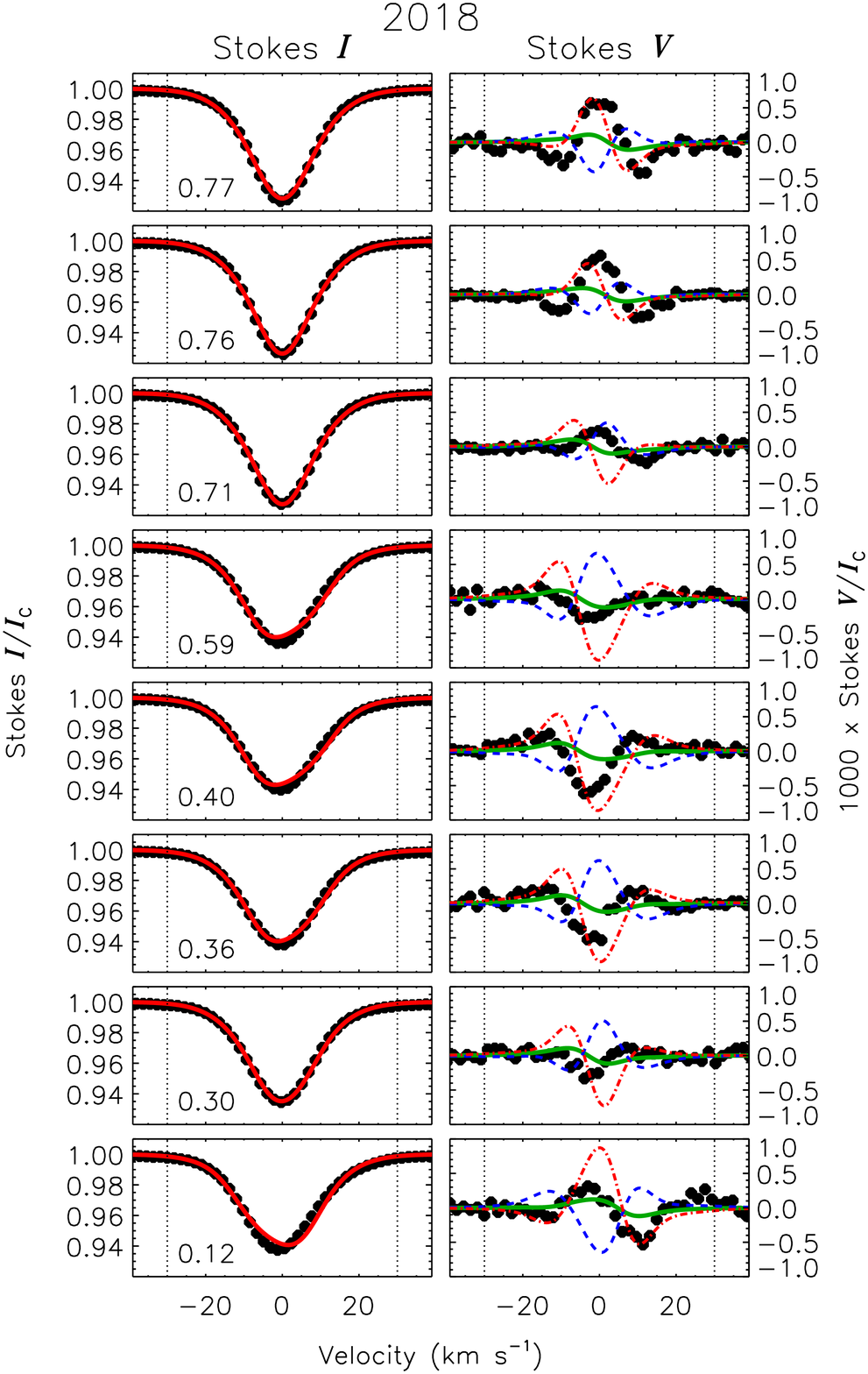} &
   \includegraphics[width=9cm]{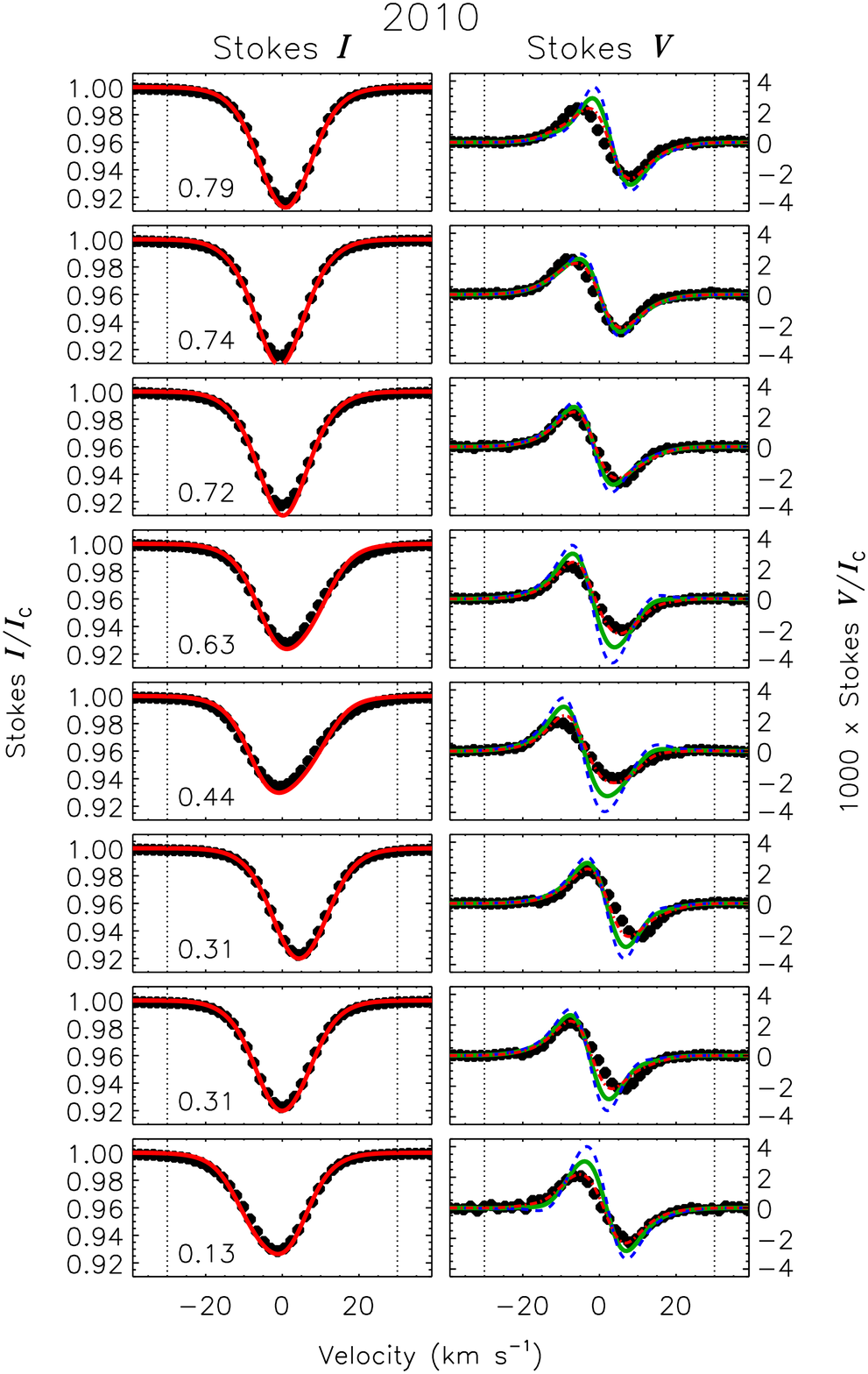} \\
	\end{tabular}
      \caption[]{{\em Left panels}: Stokes $I$ (left column) and Stokes $V$ (right) LSD profiles from 2018, shifted to their central velocities. Pulsation phase is indicated in the bottom left of each row. Observations are shown by black circles. Vertical dotted lines show the integration ranges for \bz. The model fit to Stokes $I$ is indicated by a red line. Stokes $V$ models correspond to the dipole and dipole+quadrupole models shown in Fig.\ \ref{xi1cma_bz_bq}, with identical colours and line styles. The dipolar model does not predict a crossover signature in 2018; predicts a lower Stokes $V$ amplitude than is observed; and does not reverse sign with pulsation phase. Both of the dipole+quadrupole models exhibit crossover signatures and pulsational sign reversal. Of the two, the model with $B_{\rm q} = -1$~kG provides the closest match at most phases. {\em Right panels}: same as the left panels for the 2010 observations (at \bz$_{\rm max}$) with the pulsation phases closest to those of the 2018 data. Note that different vertical scales of the Stokes $V$ panels, due to the much higher Stokes $V$ amplitude in 2010 as compared to 2018. The behaviour of the synthetic Stokes $V$ profiles does not differ as dramatically between models as at magnetic null; nevertheless, the $B_{\rm q} = -1$~kG model provides a better fit to the data than the dipolar model.}
         \label{xi1cma_stokesv_bq_crossover}
   \end{figure*}

   \begin{figure}
   \centering
   \includegraphics[width=6cm,angle=90]{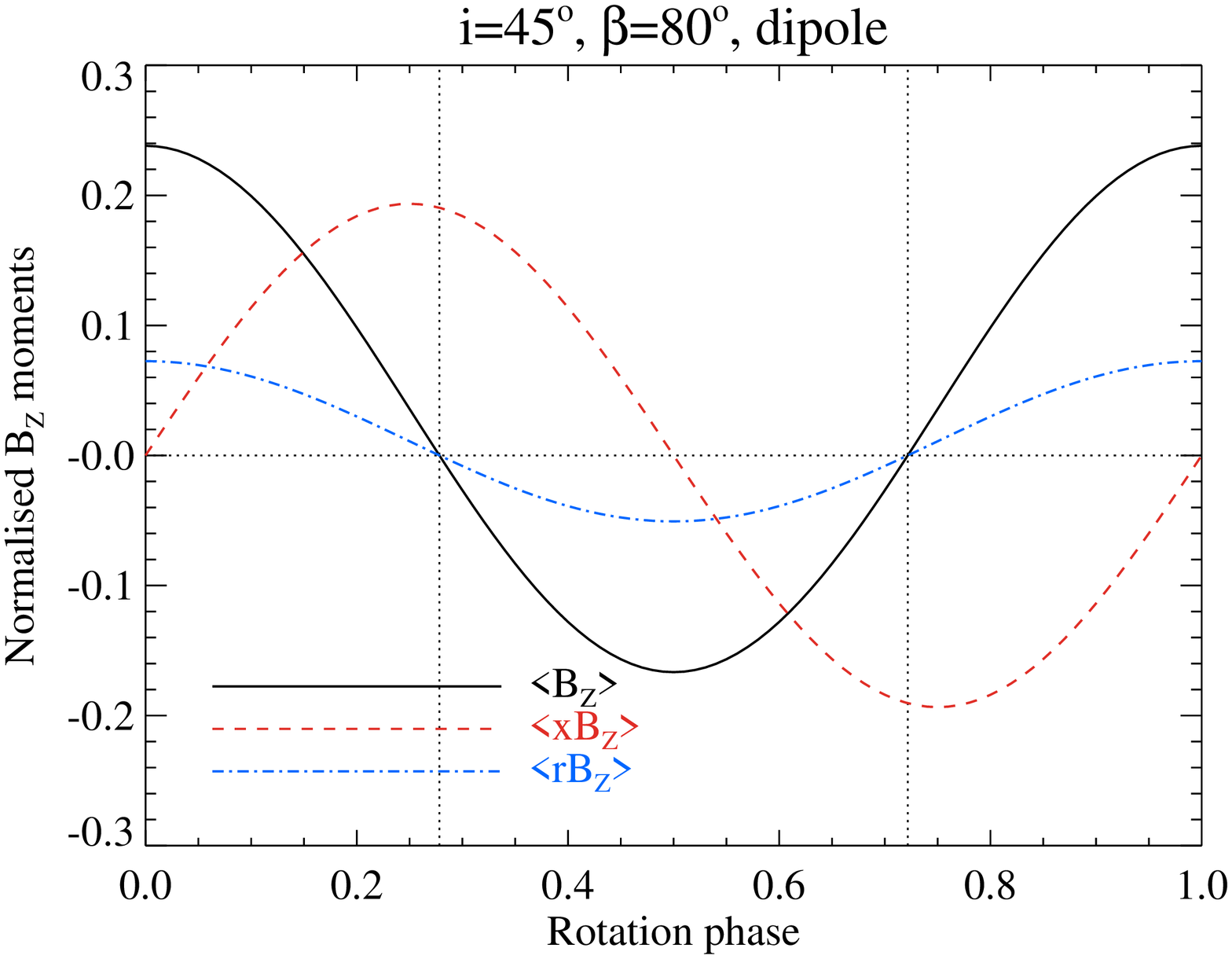} 
   \includegraphics[width=6cm,angle=90]{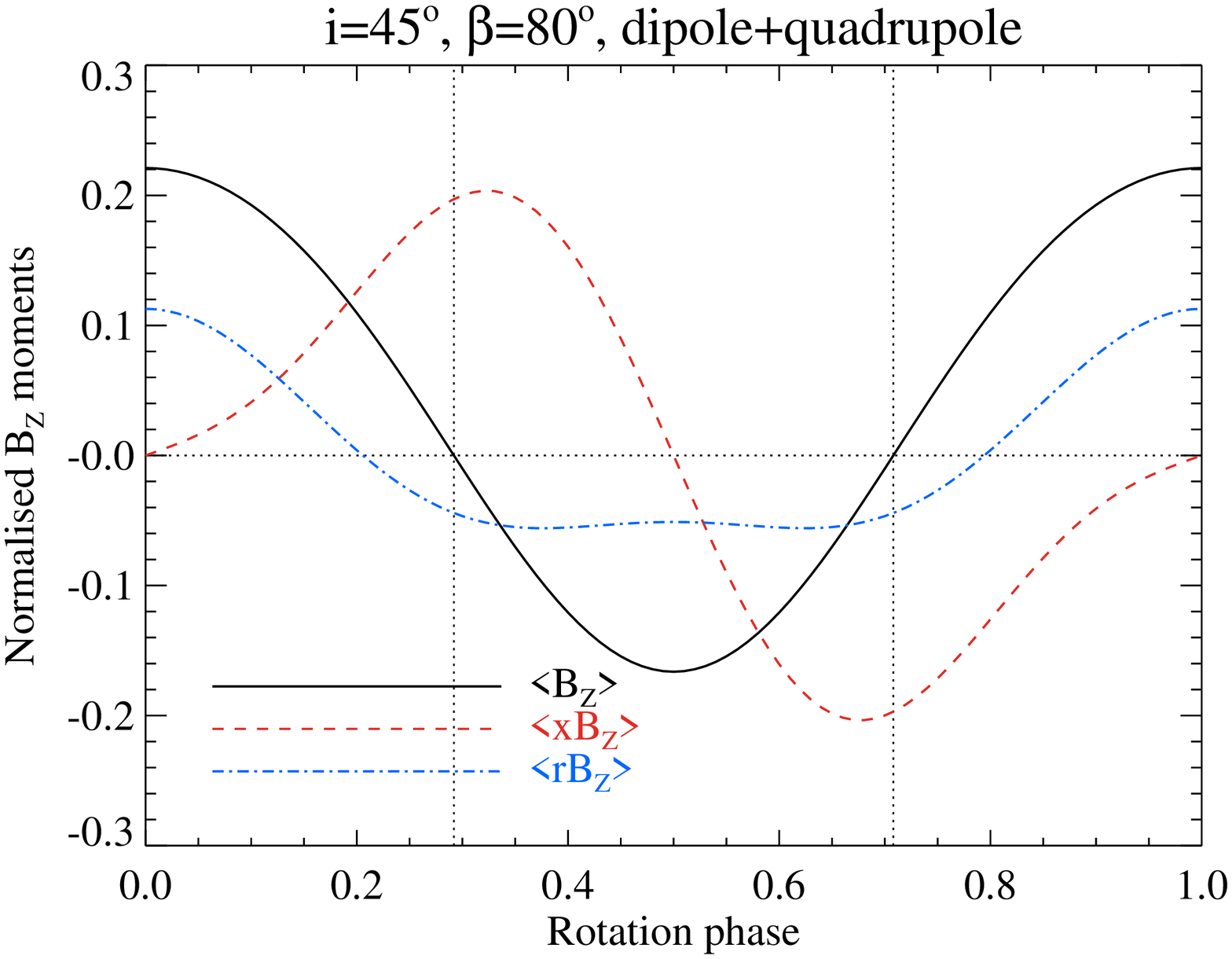} 
      \caption[]{Longitudinal magnetic field \bz, crossover $\langle x B_z \rangle$, and radial crossover $\langle r B_z \rangle$ curves for ({\em top}) a dipolar model and ({\em bottom}) a quadrupolar model with $B_{\rm q} = -B_{\rm d}$. The phases of magnetic nulls are indicated by dotted lines. The models were calculated using the same angular parameters and limb darkening constant as was used for Figs.\ \ref{xi1cma_bz_bq} and \ref{xi1cma_stokesv_bq_crossover}.}
         \label{moments}
   \end{figure}

ESPaDOnS is a fibre-fed echelle spectropolarimeter mounted at the Canada-France-Hawaii Telescope (CFHT). It has a spectral resolution $\lambda/\Delta\lambda \sim 65,000$, and a spectral range from 370 to 1050 nm over 40 spectral orders. Each observation consists of 4 polarimetric sub-exposures, between which the orientation of the instrument's Fresnel rhombs are changed, yielding 4 intensity (Stokes $I$) spectra, 1 circularly polarized (Stokes $V$) spectrum, and 2 null polarization ($N$) spectra, the latter obtained in such a way as to cancel out the intrinsic polarization of the source. \cite{2016MNRAS.456....2W} describe the reduction and analysis of ESPaDOnS data in detail. 

We obtained 8 Stokes $V$ spectropolarimetric sequences between 01/30/2018 and 02/07/2018, using the same sub-exposure time (75 s) as in the 2017 data (S17). The observation log is given in Table \ref{obs_log}. The data quality is comparable to previous seasons, with a peak signal-to-noise ratio per 1.8 \kms~pixel ranging between 579 and 933. 

Radial velocities (RVs) were measured using the same multi-line method as described by S17 (see Table \ref{obs_log}). These RVs phase coherently with the previous measurements without modification of the non-linear pulsation ephemeris determined by S17 ($P_0 = 0.2095763(1)$ d and
$\dot{P} = +0.0096(5)~{\rm s~yr^{-1}}$). Pulsation phases calculated using this ephemeris are given in Table \ref{obs_log}.

Least-Squares Deconvolution (LSD) mean line profiles were extracted using the same customized Vienna Atomic Line Database (VALD3; \citealt{piskunov1995, ryabchikova1997, kupka1999, kupka2000}) line mask used by S17. Detection flags were determined using False Alarm Probabilities \citep[FAPS; ][]{1992AA...265..669D,d1997}, with an integration range of $\pm$30~\kms, after shifting each LSD profile to its rest velocity by subtracting the measured RV. These are given in Table \ref{obs_log}. All diagnostic null $N$ LSD profiles yield a non-detection (ND; ${\rm FAP} > 10^{-3}$). All but two of the Stokes $V$ profiles are definite detections (DD; ${\rm FAP} < 10^{-5}$), with the remaining observations yielding marginal detections (MD; $10^{-5} < {\rm FAP} < 10^{-3}$). 

The longitudinal magnetic field \bz~\citep{mat1989} was also measured with an integration range of $\pm$30~\kms. \bz~measurements are given in Table 1. All \bz~and \nz~measurements are close to 0 G, with an error bar-weighted mean \bz~across all 8 measurements of $9 \pm 3$ G, a standard deviation in \bz~of 8 G, a weighted mean \nz~of $0 \pm 3$ G, and a standard deviation in \nz~of 8 G. 


\section{Results}\label{sec:results}

In Fig.\ 13 of S17, a sinusoidal fit to the \bz~measurements available at that time predicted that in 2018 the \bz~curve should pass through 0 G. This is indeed the case, as can be seen in Fig.\ \ref{xi1cma_bz_bq}, where the full set of high-resolution \bz~measurements are shown as a function of time and a coherent variation extending to the most recent year is plainly apparent. The harmonic fit presented by S17 predicts that \bz~should be $21 \pm 16$~G at the time of the new observations, which overlaps with the weighted mean \bz~of the 2018 data. 

\cite{2018arXiv180107017J} recently reported that they were unable to confirm ultra-slow rotation of $\xi^1$ CMa using low-resolution FORS2 \bz~measurements. It is likely that the inability of \cite{2018arXiv180107017J} to detect a long-term variation in \bz~is simply a consequence of the much larger error bars ($\langle \sigma_B \rangle \sim 42$~G) of their FORS2 data. However, we note that the mean of their 2017 \bz~values, 151 G, is much lower than the mean of the previous FORS1/2 measurements (290 G), indicating that a systematic decline in \bz~is indeed apparent in their data.

It is curious that the Stokes $V$ profiles yield definite detections. With $P_{\rm rot} = 30~{\rm yr}$ and the stellar radius $R_*=7.9 \pm 0.6~R_\odot$ (S17), the implied equatorial rotational velocity is $0.037 \pm 0.003$ \kms, and \vsini~should be even lower. In more rapidly rotating stars, Zeeman signatures remain detectable at \bz~$=0$ via the so-called `crossover effect' \citep{1995A&A...293..733M}. While net polarization is zero across the full line, Doppler broadening lifts flux cancellation at any given velocity. When \vsini~is essentially zero, however, this effect should disappear and the Zeeman signature should be undetectable when \bz~$\sim 0$.

Individual LSD profiles from 2018 are shown in Fig.\ \ref{xi1cma_stokesv_bq_crossover}, arranged from bottom to top in order of increasing pulsation phase and shifted to the rest velocity. For comparison, LSD profiles from 2010 with pulsation phases approximately matching those of the 2018 observations are also shown. In contrast to the 2010 data, all of which show an S-shaped Zeeman signature, crossover signatures are clearly apparent in each new observation. Curiously, the polarity of the crossover signature reverses, an effect that is most obviously apparent when comparing observations at phases $0.3-0.4$ to those obtained at phases $0.76-0.77$. Reversal of crossover signature polarity should only happen at opposite rotational phases. The modulation of the crossover signature is apparently coherent with pulsation phase over multiple pulsation cycles.

\section{Discussion}\label{sec:discussion}

Since $\xi^1$ CMa is a pure radial pulsator \citep{1994AAS..105..447H,saesen2006}, we first attempted to model Stokes $V$ by including radial pulsation using the LSD profile modelling routine described by S17. This has no effect at the magnetic null: when \bz~and \vsini~are both 0, the model predicted Stokes $V$ to be flat regardless of pulsation phase. 

We next explored the possibility that there may be contributions to $\xi^1$ CMa's surface magnetic field corresponding to spherical harmonics higher than first order. As demonstrated in Fig.\ \ref{xi1cma_bz_bq}, a quadrupolar magnetic field of $\pm 1$~kG can be added to the dipolar component ($B_{\rm d} \sim 1.4$~kG) without substantially affecting the \bz~curve. In Fig.\ \ref{xi1cma_stokesv_bq_crossover}, synthetic Stokes $V$ profiles are shown for dipolar and dipolar+quadrupolar models. We adopted a linear limb-darkening parameter $\epsilon = 0.36$, 18 \kms~of radial/tangential macroturbulent broadening, and a pulsation velocity projection factor of 1.45 (S17). For the model we assumed a 30 year rotation period, with the zero-point set in 2010, i.e. at the time of maximal \bz~(S17). In Fig.\ 2 we show model signatures for rotational phases 0.27 (in 2018) and 0.0 (in 2010). Since a rotation phase of 0.27 does not correspond exactly to a magnetic null, the dipolar model still yields a weak Stokes $V$ signature; however, the amplitude is much weaker than is observed, both the amplitude and the shape are essentially independent of pulsation phase, and it is not a crossover signature. By contrast, both of the dipole+quadrupole models yield Stokes $V$ signatures exhibiting 1) a crossover morphology; 2) pulsationally coherent variability of both sign and amplitude; 3) inversion of crossover signatures at different pulsation phases. Of the two models, the one with $B_{\rm q} = -1$~kG provides the closest match to the observed profiles. This is most apparent at pulsation phases 0.12, 0.4, and 0.77. 


For comparison, the right-hand panels of Fig.\ \ref{xi1cma_stokesv_bq_crossover} show models for the observations obtained in 2010 (i.e., at \bz$_{\rm, max}$), with pulsation phases corresponding most closely to those of the 2018 data. While the differences between models are less drastic at this rotation phase, the model with $B_{\rm q}=-1$ kG still gives the best match to observations, both in terms of amplitude and the rounded wings of the Stokes $V$ profile, which are reproduced by the negative quadrupolar model but not by the positive or pure dipolar models. 

This phenomenon can be explained by a `radial crossover' effect. The crossover $\langle x B_ z \rangle$ is measured via the second moment of Stokes $V$ \citep{1995A&A...293..733M} weighted by the projected surface area and a linear continuum limb darkening function, where the $z$ coordinate is aligned with the line of sight and the $x$ coordinate is in the plane of the sky and perpendicular to the rotational axis. For a rotating star isovelocity contours are parallel to the rotational axis (i.e. in the $y$ direction). As demonstrated in Fig.\ \ref{moments}, for a dipole $\langle x B_{\rm z} \rangle$ is a maximum at \bz~$=0$, and is also non-zero at \bz~$=0$ for a dipole+quadrupole. In all cases it should disappear entirely when \vsini~$=0$, as all surface elements share the same isovelocity contour (i.e. $\Delta v=0$~\kms). Motivated by the radial symmetry of the pulsational velocity field we define the {\em radial crossover} as $\langle r B_{\rm z} \rangle$ where $r = \sqrt{x^2 + y^2}$. In this case isovelocity contours share the same $r$ coordinate. Radial crossover curves are shown in Fig.\ \ref{moments}. For a centred dipole $\langle r B_{\rm z} \rangle = 0$ at magnetic nulls due to the meridional symmetry of this configuration, since the integral of $B_z$ along a given isovelocity contour is still 0. However, for a dipole+quadrupole $\langle r B_{\rm z} \rangle \ne 0$ at \bz~$=0$ because, for this configuration, the meridional symmetry of the magnetic field is broken, and the integral of $B_z$ along a given isovelocity contour is no longer 0. Thus, radial pulsations are able to induce detectable crossover signatures for magnetic fields more complex than centred dipoles. 

While Fig.\ \ref{moments} shows moment phase curves calculated using the same angular parameters as adopted in Figs.\ \ref{xi1cma_bz_bq} and \ref{xi1cma_stokesv_bq_crossover}, we examined models for a range of angular parameters and found that the fundamental result, that $\langle r B_{\rm z} \rangle \ne 0$ at \bz~nulls for magnetic fields more complex than a dipole, is independent of $i$ and $\beta$. We further emphasize that the Stokes $V$ models shown in Fig.\ \ref{xi1cma_stokesv_bq_crossover} are not meant as detailed fits to the observations, but as a demonstration that a radial crossover model provids a good qualitative description of the (otherwise unexplained) general modulation of shape and amplitude of the observed Stokes $V$ signatures across a pulsation cycle.

S17 reported a possible pulsational modulation of \bz, in the form of a weak ($\pm15$ G) variation with the pulsation period superimposed on the dominant rotational modulation, with an amplitude correlated to the magnitude of \bz. The effect reported here does not lead to a pulsational modulation of \bz~at any rotational phase, therefore the reported modulation, if real, must be related to some other effect, e.g. flux conservation in a stellar atmosphere with a changing radius. Verification will require high-cadence sampling of the magnetic field curve with a precision of a few G, and this is furthermore best performed at a \bz~extremum. The next such opportunity is likely to occur in approximately 2025. 

\section{Conclusions}\label{sec:conclusions}

High-resolution ESPaDOnS magnetic measurements obtained in 2018 confirm the extremely slow rotation of $\xi^1$ CMa proposed by S17. In accordance with the predictions of a harmomic fit to the \bz~measurements obtained between 2000 and 2017, \bz~is approximately 0 G in 2018. We have also obtained new RV measurements, which can be coherently phased with previous data using the non-linear pulsation ephemeris determined by S17 without modification.

All LSD Stokes $V$ profiles display a Zeeman signature closely resembling the classic magnetic crossover expected at a magnetic null, but {\em not} expected when \vsini~$\sim 0$ as is the case for $\xi^1$ CMa. These crossover signatures exhibit a pulsationally coherent variation, with large changes in amplitude as well as apparent inversion of sign. We have shown that these phenomena can be explained by means of a `radial crossover' signature induced by the domination of the star's velocity field by its radial pulsations, in conjunction with a surface magnetic topology differing from a centred dipole. 

We do not claim that the magnetic field of $\xi^1$ CMa is accurately described by anti-aligned dipolar and quadrupolar components, indeed, cursory examination of Fig.\ \ref{xi1cma_stokesv_bq_crossover} shows that the fit achieved using an aligned quadrupole is only approximate. It is more likely that the star's magnetic topology is a distorted dipole. However, this effect opens the intriguing possibility that pulsation might be exploited to probe the detailed surface magnetic topologies of magnetic pulsating stars. The pulsational modulation of the circular polarization profiles of other such objects, such as $\beta$ Cep \citep{henrichs2013}, should be re-examined with this possibility in mind. 

The magnetic model presented here and the 30-year rotational period predict that in 2019 the star's Zeeman signature should be negative for the first time since it was first observed with MuSiCoS in 2000, with an amplitude of about $-35$~G. While this is slightly further from the \bz~null than the current year, the model described here predicts that the radial crossover phenomenon may still be detectable in 2019. 

\section*{Acknowledgements}

This work has made use of the VALD database, operated at Uppsala University, the Institute of Astronomy RAS in Moscow, and the University of Vienna. MS is supported by Natural Sciences and Engineering Research Council (NSERC) of Canada Postdoctoral Fellowship program. GAW acknowledges support from a Discovery Grant from NSERC. OK acknowledges financial support from the Knut and Alice Wallenberg Foundation, the Swedish Research Council, and the Swedish National Space Board. The authors acknowledge Nadine Manset for her verification that nothing unusual was affecting ESPaDOnS during the time in which the observations were collected.

\bibliography{bib_dat.bib}{}


\end{document}